\documentclass[aps,11pt]{revtex4}
\usepackage[colorlinks=true,linkcolor=blue,urlcolor=blue,filecolor=black,citecolor=red,pdfstartview=FitV,pdftitle={},pdfsubject={},
pdfkeywords={},pdfpagemode=None,bookmarksopen=true]{hyperref}
\usepackage{graphicx}%Include figure filespp
\usepackage{epstopdf}%
\usepackage{amsmath}
\usepackage{amsfonts}
\usepackage{amssymb}
\usepackage{color}%
\usepackage{dcolumn}% Align table columns on decimal pointppp
\usepackage{slashed}
\usepackage{amssymb,ulem}
\usepackage{float}
\usepackage{amsthm,amsmath,amssymb}
\usepackage{mathrsfs}
\usepackage{subfigure}
\usepackage{wrapfig}
\usepackage{indentfirst}
\usepackage{multirow}
\usepackage{appendix}
\makeatletter

\newcommand{\Rmnum}[1]{\expandafter\@slowromancap\romannumeral #1@}

\makeatother

\begin{document}
\title{Spontaneous scalarization of Bardeen black holes}
\author{Lina Zhang$^{1,2}$\footnote{linazhang@hunnu.edu.cn;},  Qiyuan Pan$^{1,2}$\footnote{Corresponding author: panqiyuan@hunnu.edu.cn;}, Yun Soo Myung$^{3}$\footnote{ysmyung@inje.ac.kr;}, De-Cheng Zou$^{4}$\footnote{Corresponding author: dczou@jxnu.edu.cn}}

\affiliation{$^{1}$Key Laboratory of Low Dimensional Quantum Structures and Quantum Control of Ministry of Education, Synergetic Innovation Center for Quantum Effects and Applications, and Department of Physics, Hunan Normal University, Changsha, Hunan 410081, China\\
$^{2}$Institute of Interdisciplinary Studies, Hunan Normal University, Changsha, Hunan 410081, China\\
$^{3}$Institute of Basic Sciences and Department of Computer Simulation, Inje University, Gimhae 50834, Republic of Korea\\
$^{4}$College of Physics and Communication Electronics, Jiangxi Normal  University, Nanchang 330022, China 
}

\date{\today}

\begin{abstract}
\indent
We study the spontaneous scalarization of Bardeen black holes, whose tachyonic instability triggers the formation of scalarized charged black holes (SCBHs). In this case, we find infinite  ($n=0,1,2,\cdots$) branches  of  SCBHs with magnetic charge $g$. The $n = 0$ branch of SCBHs can be found for the coupling parameter $\alpha \geq \alpha_{n=0}(g)$ with both quadratic (1-$\alpha \varphi^2$) and exponential ($e^{-\alpha \varphi^2}$) couplings, where $\alpha_{n=0}(g)$ represents the threshold of tachyonic instability for the Bardeen black holes.
Furthermore, it is shown that the $n = 0$ branch for  both couplings is stable against radial perturbations.
This stability shows that this branch can be used for further observational implications.

\end{abstract}

\maketitle

%%%%%%%%%%%%%%%%%%%%%%%%%%%%%%%%%%%%%%%%%%%%%%%%%%%%%%%%%%%%%%%%%%%%%%%%%%%
\section{Introduction}\label{1s}

Spontaneous scalarization is a dynamic process that imparts scalar hair to black holes (and other compact objects) without changing the predictions in the weak field limit \cite{Damour:1993,Damour:1996,Doneva:2018,Silva:2018,Antoniou:2018}. This phenomenon is a strong gravity phase transition caused by tachyonic instability resulting from the nonminimal coupling between scalar fields and spacetime curvature or matter.
Black hole spontaneous scalarization has been extensively studied \cite{Doneva:2018,Silva:2018,Antoniou:2018,Minamitsuji:2019,Silva:2019,Doneva:2019,Macedo:2019,Blázquez-Salcedo:2022,Antoniou:2021}, including cases involving rotation \cite{Cunha:2019,Collodel:2020} and spin-induced scalarization \cite{Dima:2020,DonevaPRD:2020,DonevaEPJC:2020,Herdeiro:2021,Berti:2021}. These black holes are found to be entropically favorable compared to bald (general relativity) solutions and their $n=0$ branches are stable \cite{Blázquez-Salcedo:2018,Blázquez-Salcedo101:2020,Blázquez-Salcedo102:2020}. The nonlinear dynamics of scalarized black holes in  scalar-Gauss-Bonnet(sGB) gravity, including mergers and stellar core collapse, have been examined \cite{Ripley:2020,Silva:2021,Doneva064024:2021,Kuan:2021,East:2021}. Additionally, spontaneous scalarization has been explored in other alternative theories of gravity \cite{Herdeiro:2018,Andreou:2020,Gao:2019,Doneva:2021,Zhang:2021,Myung:2021}. These includes the Einstein-Maxwell-scalar theory with exponential and quadratic scalar couplings \cite{Myung:2018vug,Myung:2018jvi}.

In general relativity, singularity theorems \cite{Hawking:1973} suggest that singularities are inevitable inside black holes. It is worth noting that  these are considered nonphysical and  may be avoided in an alternative theories of gravity. In this context, Bardeen \cite{Bardeen:1968} proposed the first regular black hole solution, which is spherically symmetric and free of singularities.
The physical source of Bardeen black holes was initially unclear. By the end of the last century, nonlinear electromagnetic sources were proposed to explain the matter content \cite{Ayon-Beato:1998,Ayon-Beato:2000}, suggesting that regular black holes could be obtained due to nonlinear electric charge or magnetic monopoles. Other similar solutions  were also found when using nonlinear electrodynamics \cite{Hayward:2006,Kumara:2023,Fan:2016,Berej:2006}. We note that  regular black holes are of great interest for understanding fundamental issues in physics, including singularities and nonlinear electrodynamics \cite{Huang:2020,Zou:2021}.
In this work, hence,  we would like to study the spontaneous scalarization of Bardeen black holes by introducing two scalar field couplings.

The work is organized as follows. In Sec.~\ref{2s}, we  introduce the  Einstein-nonlinear electrodynamics theory coupled with scalar field.
Sec.~\ref{3s} is devoted to discuss the tachyonic instability of the Bardeen black holes.
In Sec.~\ref{4s}, we consider two scalar field coupling forms to  derive the $n = 0$ branch of SCBHs numerically.
We wish to analyze the stability of $n = 0$ branch of  SCBHs in Sec.~\ref{5s}.
Finally, we close the work with discussions and conclusions in Sec.~\ref{6s}.

\section{The Theoretical Framework}
\label{2s}
We consider Einstein-nonlinear electrodynamics theory with scalar coupling function described by the
following action functional
\begin{equation}\label{eq4:action}
  I=\frac{1}{16 \pi} \int d^{4} x \sqrt{-g}
  \Big[\mathcal {R}  -2(\nabla \varphi)^2 -4 \tilde{f}(\varphi) \mathcal{L}(\mathcal{F}) \Big],
\end{equation}
where $\mathcal {R}$ is the scalar curvature, $\varphi$ is the scalar field and a coupling function $\tilde{f}(\varphi)$ depending on $\tilde{f}(\varphi)$.
Further, $\mathcal{L}(\mathcal{F})$ is a nonlinear  function of $\mathcal{F}=F^2 = F_{\mu\nu} F^{\mu\nu}$ with
 $F_{\mu\nu} = \partial_\mu A_\nu - \partial_\nu A_\mu$ defined by
\begin{equation}\label{eq2:LF}
\mathcal{L}(\mathcal{F}) = \frac{3}{2 s g^2}  \left(\frac{\sqrt{2g^2 F^2/2}}{ 1 + \sqrt{2g^2 F^2/2} }\right)^{\frac{5}{2}},
\end{equation}
where the parameter $s$ is given by $s = \frac{|g|}{2M}$, $g$ and $M$ are free parameters associated with the magnetic charge and  mass, respectively.

Varying the action with respect to $g_{\mu\nu}$, $\varphi$, and $A_\mu$ gives three field equations 
\begin{eqnarray}
  G_{\mu\nu}&\equiv& 2T_{\mu\nu}=
  2\tilde{f}(\varphi)\Big[4\frac{\partial \mathcal{L}(\mathcal{F})}{\partial \mathcal{F}} F_{\mu\eta}F^{\eta}_{\nu}-g_{\mu\nu}\mathcal{L}(\mathcal{F}) \Big]
  +2\partial_{\mu}\varphi\partial_{\nu}\varphi  - (\nabla\varphi)^2 g_{\mu\nu}, \label{eq4:eist} \\
 && \nabla^2\varphi =\frac{\partial \tilde{f}(\varphi)}{\partial \varphi} \mathcal{L}(\mathcal{F}),\label{eq4:dilaton}\\
  &&\nabla_{\mu}\Big[4\tilde{f}(\varphi) \frac{\partial \mathcal{L}(\mathcal{F})}{\partial \mathcal{F}}F^{\lambda\mu}\Big]  = 0.\label{eq4:magnetic}
\end{eqnarray}

Taking into account $\varphi = 0$, the Bardeen black hole solution is obtained by solving  Eqs.(\ref{eq4:eist})(\ref{eq4:magnetic}) \cite{Ayon-Beato:1998,Ayon-Beato:2000}
\begin{eqnarray}\label{eq4:metric}
  ds_{\text{Bardeen}}^2 = -f(r)  dt^2 +\frac{dr^2 }{f(r)}+ r^2 d\theta^2 + r^2 \sin^2\theta \, d\phi^2,
\end{eqnarray}
with the metric function
\begin{eqnarray}\label{Bardeen-solution}
f(r) = 1 - \frac{2Mr^2}{\left(r^2 + g^2\right)^{\frac{3}{2}}}.
\end{eqnarray}
Here $g$ and $M$ are the magnetic charge and  mass of Bardeen black hole, respectively.
In this case, the magnetic field strength is expressed  as 
\begin{eqnarray}\label{Bardeen solution}
F_{\mu\nu} = 2\delta_{[\mu}^{\theta} \delta_{\nu]}^{\phi} g \sin\theta,
\end{eqnarray}
where  we have $F_{\theta\phi} = g \sin\theta(A_\phi=-g\cos \theta)$ and $F^2= \frac{2g^2}{r^4}$.
In this case, computing the energy-momentum tensor $T_{\mu}^{~\nu}={\rm diag}[-\rho,p_r,p_t,p_t]$, there is the violation of strong energy condition ($\rho+p_r+2p_t<0$) at the center, implying the regular black hole \cite{Ayon-Beato:1998,Ayon-Beato:2000}. 

\section{Instability for Bardeen black holes}\label{3s}

We briefly mention the tachyonic instability of Bardeen black hole as it serves as the starting point for spontaneous scalarization. In this paper, we choose two coupling forms: $\tilde{f}(\varphi) = 1 - \alpha \varphi^2$, representing a quadratic coupling with parameter $\alpha$ and $\tilde{f}(\varphi) = e^{-\alpha \varphi^2}$, denoting an exponential coupling.
Based on the Klein-Gordon equation \eqref{eq4:dilaton},
the linearized equation for a perturbed scalar $\delta\varphi$ is expressed as
\begin{eqnarray}\label{perturbed-scalar}
\bar{\nabla}^2 \delta\varphi + 2 \alpha \mathcal{L}(\mathcal{F}) \delta\varphi = 0,
\end{eqnarray}
which determines the tachyonic instability of  Bardeen black hole.
The last term in (\ref{perturbed-scalar}) represents an effective mass term, leading to the instability of  Bardeen black hole which is contingent on the coupling parameter $\alpha$.
Considering $M = 0.5$ and $g = 0.2$ as a typical nonextremal Bardeen black hole, one can yield an outer horizon $r=r_+ = 0.935$ from $f(r) = 0$ in Eq.~\eqref{Bardeen-solution}, for example.

\begin{figure}[H]
  \subfigure[]{\label{fig:1:1} %% label for second subfigure
  \includegraphics[width=8cm]{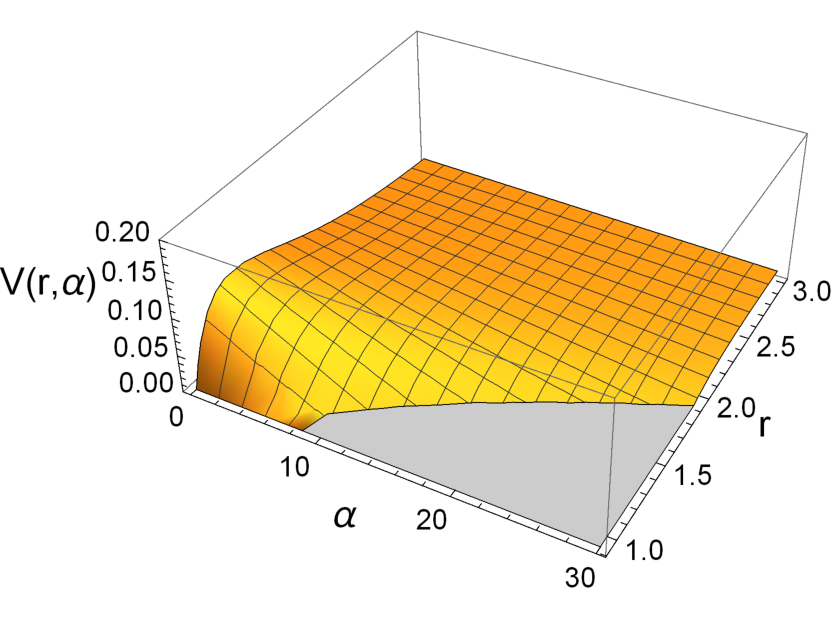}}%\\
   \quad\quad
  \subfigure[]{\label{fig:1:2} %% label for second subfigure
  \includegraphics[width=8cm]{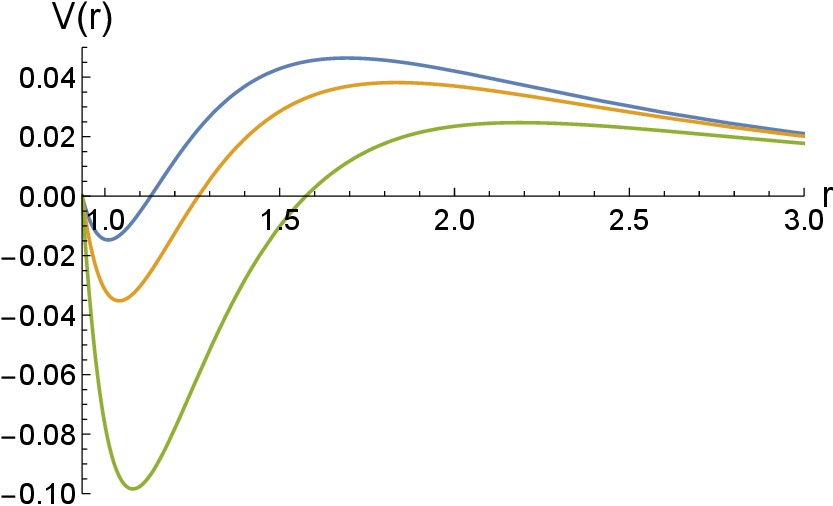}}%\\
  \caption{(a) The $\alpha$-dependent potential $V(r, \alpha, g=0.2)$ as a function of $r \in [r_+, 3.0]$ and $\alpha \in [0.01, 30]$ for the outer horizon radius $r_+=0.935 (M=0.5, g=0.2)$. The shaded region along the $\alpha$-axis represents the negative region of the potential. (b) Plots of potentials $V(r, \alpha, g=0.2)$ with three different values $\alpha = \{10, \alpha_\text{th} = 12.712, 20\}$ from top to bottom near the $V$-axis.}
    \label{fig:1}
\end{figure}
Now, we use the separation of variables for the spherically symmetric Bardeen background (\ref{eq4:metric}) given by
\begin{equation}
\varphi(t, r, \theta, \phi) = \frac{u(r)}{r} e^{-i\omega t} Y_{lm}(\theta, \phi). \label{11}
\end{equation}
Choosing a tortoise coordinate $r_*$, defined by $r_* = \int \frac{dr}{f(r)}$, we obtain the radial part of the scalar equation as
\begin{equation}\label{ur}
\frac{d^2 u}{dr_*^2} + \left[\omega^2 - V(r)\right] u(r) = 0.
\end{equation}
Here the scalar potential $V(r)$ is expressed as
\begin{equation}\label{Vr}
V(r)= f(r)\Bigg[\frac{l(l+1)}{r^2} + \frac{2 M \left[r^2 - g^2 \left(2 + 3 \alpha \right)\right]}{\left(g^2 + r^2\right)^{5/2}}\Bigg].
\end{equation}
 The $s$($l = 0$)-mode is permissible for scalar perturbations and can therefore be used to assess the instability of  Bardeen black hole. From now on, we will focus on the $l = 0$ mode.
From the potential (\ref{Vr}), the sufficient condition for stability requires that the potential be positive definite outside the event horizon, expressed as $V(r) \geq 0$ \cite{Myung:2019}.
However, deriving the instability condition from potential (\ref{Vr}) is challenging, so we observe the negative region near the horizon as a signal of instability.
We show the negative region of potential (\ref{Vr}) as a function of $r$ and $\alpha$ in Fig. \ref{fig:1:1}. Fig. \ref{fig:1:2} indicates that the width and depth of the negative region in $V(r, \alpha)$ increase with $\alpha$. 
If the potential $V(r)$ is negative in the near-horizon, it is conjectured that this may lead to a growing perturbation in the spectrum, indicating tachyonic instability of a Bardeen black hole.
However, this is not always true.
\begin{figure}[H]
  \centering
  \includegraphics[width=8cm]{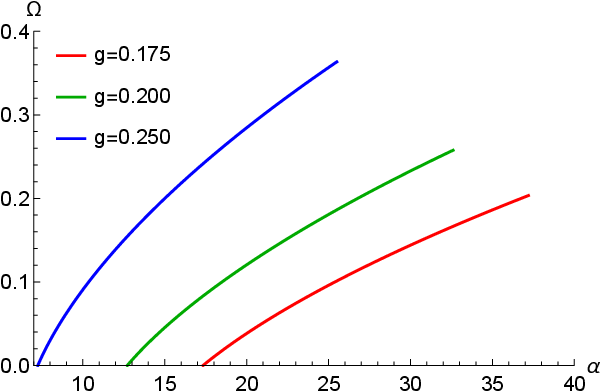}%\\
  \caption{Three curves of $\Omega$ in $e^{\Omega t}$ as a function of $\alpha$ are used to determine the thresholds of instability $[\alpha_\text{th}(g)]$ around a Bardeen black hole. We find $\alpha_\text{th}(g) = 17.338(0.175), 12.712(0.200), 7.251(0.250)$ when three curves cross $\alpha$-axis.}
  \label{fig:2}
\end{figure}
A key factor in determining the stability of a black hole is whether the scalar perturbation decays over time.
The linearized scalar equation (\ref{ur}) around a Bardeen black hole  permits an unstable (growing) mode such as $e^{\Omega t}$ for scalar perturbations, signaling instability in the black hole.  Notably, this instability often leads to the emergence of scalarized black holes.
Therefore, we solve equation (\ref{ur}) numerically after substituting $\omega = -i\Omega$, by imposing boundary conditions of a purely ingoing wave at the near-horizon and a purely outgoing wave at infinity. From Fig. \ref{fig:2}, we read off the threshold of instability $[\alpha_\text{th}(g)]$.
Thus, the instability bound can be determined numerically by
\begin{equation}\label{eq:instability_bound}
\alpha \geq \alpha_{\text{th}}(g),
\end{equation}
with $\alpha_\text{th}(g) = {17.338(0.175), 12.712(0.200), 7.251(0.250)}$.
On the other hand, stable Bardeen black holes exist for $\alpha < \alpha_{\text{th}}(g)$. For $g=0.2$, Fig.~\ref{fig:1:2} shows that $\alpha < \alpha_{\text{th}} = 12.712$ corresponds to stable Bardeen black holes, while $\alpha \geq \alpha_{\text{th}}$ corresponds to unstable Bardeen black holes.

\begin{figure}[H]
  \centering
  \includegraphics[width=8cm]{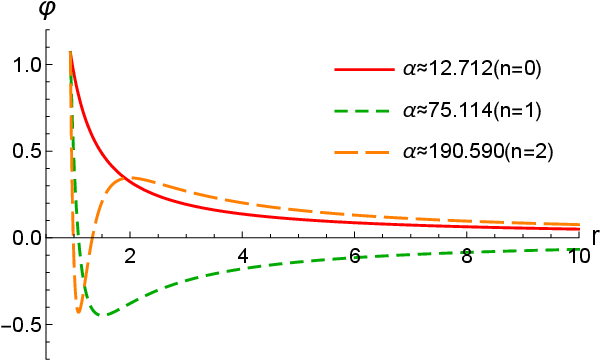}%\\
  \caption{Plot of radial profiles $\varphi(z) = u(z)/z$ as a function of $z = r/2M$ for $M = 0.5$ and $g = 0.2$, showing the first three static perturbed scalar solutions. The number $n$ of zero nodes describes the $n = 0, 1, 2$ SCBHs.}
  \label{fig:3}
\end{figure}

To check the instability bound (\ref{eq:instability_bound}), we need to precisely determine $\alpha_{\text{th}}(g)$, as it influences the formation of scalarized black holes. This can be verified by solving for a static scalar solution [scalar cloud: $\varphi(r)$] to the linearized equation (\ref{ur}) with $u(r) = r \varphi(r)$ and $\omega = 0$ in the Bardeen background.
For $l = 0$, $M = 0.5$, and $g = 0.2$, requiring an asymptotically normalizable solution yield a discrete set for $\alpha_n(g)$, where $n = 0, 1, 2, \cdots$ denotes the number of zero crossings of $\varphi(r)$ (or order number). 
See Fig.~\ref{fig:3} for static scalar solutions $\varphi(z)$ with $z = r/2M$, $M = 0.5$, and $g = 0.2$. The $n = 0$ scalar mode represents the fundamental branch of scalarized black holes, while  the $n = 1, 2$ scalar modes indicate other branches. Actually, infinite ($n=0,1,2,\cdots$) branches of SCBHs appear from  infinite scalar modes. This is  a key result for spontaneous scalarization.
We note that $\{\alpha_0,\alpha_1,\alpha_2\}$
 correspond to the first three bifurcation points for emerging the $n=0,1,2$ branches.
As is shown in Fig.~\ref{fig:2}, we confirm that for given $g=0.2$,
\begin{equation}
\alpha_{\text{th}}(g) = \alpha_{n=0}(g), 
\end{equation} 
which means that the instability threshold for Bardeen black holes means a formation of the largest $n = 0$ branch of SCBHs.

\section{Scalarized charged black holes}\label{4s}

All scalarized charged black holes will be generated
from the onset of scalarization ${\varphi_n(r)}$ in the unstable region of Bardeen black hole $[\alpha(g) \geq \alpha_{\text{th}}(g)]$. 
In order to find scalarized charged black holes numerically, one proposes the metric ansatz and fields
\begin{eqnarray}\label{eq4:metric-1}
&&  ds_{\text{SBH}}^2 = -N(r)e^{-2\delta(r)}  dt^2 +\frac{dr^2 }{N(r)}+ r^2 (d\theta^2+\sin^2\theta d\phi^2),\nonumber\\
&&  \varphi=\varphi(r)\neq 0, \quad A_{\phi}=A_{\phi}(r).
\end{eqnarray}
in which $N(r) = 1 - 2m(r)/r$, and $\delta(r)$ is the function of $r$.

Substituting the metric ansatz and fields \eqref{eq4:metric-1} into Maxwell equation (\ref{eq4:magnetic}),
we can obtain a vector potential solution $A_\phi=-g\cos \theta$, namely the magnetic field solution of $F_{\theta\phi}= g \sin\theta$ and $F^2 = \frac{2g^2}{r^4}$ like Bardeen black hole solution.
This implies that we do not need to have an approximate solution for $A_\phi$. 

We mention again that $n=0$ branch of SCBHs appears  for  $\alpha(g) \geq \alpha_{\text{th}}(g)$.  In particular, we consider two coupling forms: $\tilde{f}(\varphi) = 1 - \alpha \varphi^2$ and $\tilde{f}(\varphi) = e^{-\alpha \varphi^2}$. Using these forms, we construct the $n = 0$ branch of SCBHs numerically for $M = 0.5$ and $g = 0.2$. Similarly, we may construct other branches of SCBHs. 

Now, we introduce  the scalar $\varphi(r)$.
Plugging the metric ansatz and fields \eqref{eq4:metric-1}  into Eqs.~(\ref{eq4:eist})–(\ref{eq4:dilaton}) results in three equations for $\{ \delta(r), m(r), \varphi(r) \}$ as
\begin{eqnarray}
&&\delta'(r) + r \varphi'^2(r)=0,\\
&&\frac{6 g^2 M r^2\tilde{f}(\varphi)}{(g^2 + r^2)^{5/2}} + r(r - 2m) \varphi'^2(r) - 2 m'(r)  =0,\\
&&r ( r - 2m) \varphi''(r)-\Big \{m [ 2 - 2r \delta'(r)] + r [ 2m'(r) + r \delta'(r)-2  ] \Big\} \varphi'(r) - \frac{3 g^2 M r^2\tilde{f}'(\varphi)}{(g^2 + r^2)^{5/2}} =0,
\end{eqnarray}
where the prime ($'$) indicates differentiation with respect to the argument.
An approximate solution in the near-horizon is
\begin{align}
m(r) &= \frac{r_+}{2} + m_1 (r - r_+) + \cdots,\label{eq-1} \\
\delta(r) &= \delta_0 + \delta_1 (r - r_+) + \cdots,\label{eq-2} \\
\varphi(r) &= \varphi_0 + \varphi_1 (r - r_+) + \cdots,\label{eq-3}
\end{align}
where three coefficients are determined  by
\begin{align}
m_{1} &= \frac{3 g^2 M r_+^2 \tilde{f}(\varphi_0)}{(g^2 + r_+^2)^{5/2}},\quad\quad\delta_{1} = -r_{+} \varphi_{1}^{2},\nonumber\\ 
\varphi_{1} &= \frac{
3 g^2 M r_+ \left[ \left( g^2 + r_+^2 \right)^{5/2} + 6 g^2 M r_+^2 \tilde{f}(\varphi_0) \right] \tilde{f}' (\varphi_0)
}{\left( g^2 + r_+^2 \right)^5 - 36 g^4 M^2 r_+^4 \tilde{f}(\varphi_0)^2}.
\end{align}
The near-horizon solution involves two parameters, $\varphi_0=\varphi(r_{+}, \alpha)$ and $\delta_0=\delta(r_{+}, \alpha)$, which are determined by matching (\ref{eq-1})–(\ref{eq-3}) with the asymptotic solution in the far-region
\begin{align}
m(r) = M -  \frac{Q_s^2}{2r} + \cdots,\quad\quad
\varphi(r) = \frac{Q_s}{r} + \cdots, \quad\quad
\delta(r) = \frac{Q_s^2}{2r^2} + \cdots,
\end{align}
which incorporates the Arnowitt-Deser-Misner mass $M$ and the scalar charge $Q_s$.

\begin{figure}[H]
\centering
\subfigure[Quadratic coupling: $\tilde{f}(\varphi) = 1 - \alpha \varphi^2$ ]{
\label{fig:4}
\includegraphics[width=7cm]{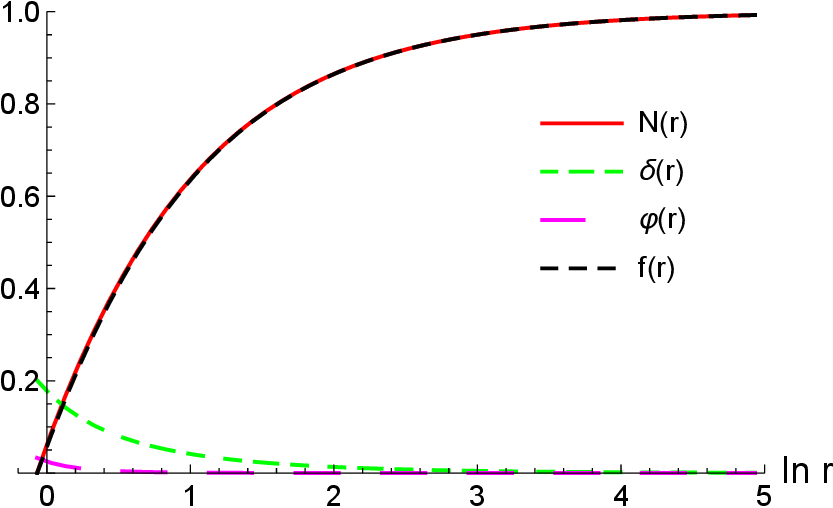}}
\hfill%
\subfigure[Exponential coupling: $\tilde{f}(\varphi) = e^{-\alpha \varphi^2}$]{
\label{fig:5}
\includegraphics[width=7cm]{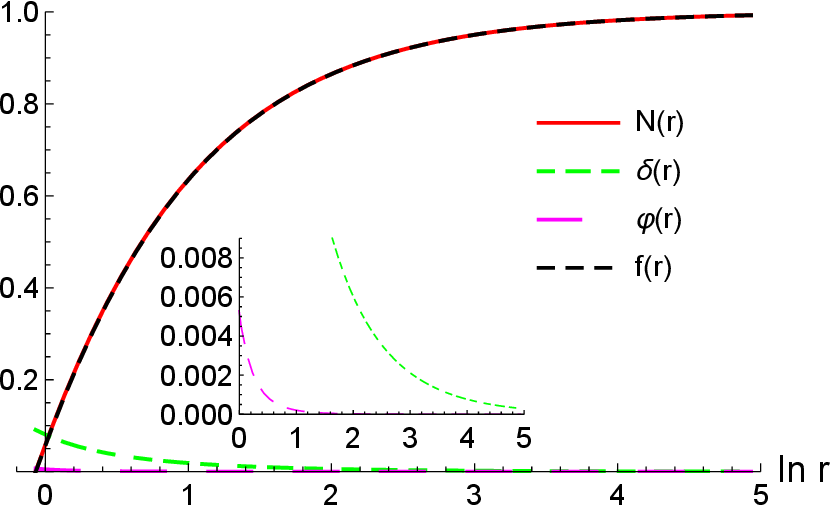}}
\caption{Plots of a SCBH solution with $g = 0.2$,  and $M=0.5$ for $\alpha = 13.048$ (Quadratic coupling) and $\alpha = 13.667$ (Exponential coupling) in the $n = 0$ branch of $\alpha \geq 12.712$. 
It shows metric functions $\delta(r)$, $N(r)$, and $f(r)$ for the Bardeen black hole, and scalar hair $\varphi(r)$. We note that metric function $N(r)$ has a horizon at $\ln r = -0.0683$ while $f(r)$ for Bardeen black hole takes a horizon at $\ln r = -0.0671$. }
\end{figure}

Consequently, for quadratic coupling, we obtain the $n = 0$ branch of SCBH solution  shown in Fig.~\ref{fig:4} for $\alpha = 13.048$ at $g = 0.2$.
The metric function $N(r)$ has a slightly different horizon at $\ln r =  -0.0683$ compared to the Bardeen horizon at $\ln r = -0.0671$, but it nearly coincides with the Bardeen metric function $f(r)$ as $\ln r$ increases.
Also, $\delta(r)$ decreases as $\ln r$ increases, while $\delta_{\text{Bardeen}}(r) $ remains zero because $e^{-2\delta(r)} = 1$ for the Bardeen case.
Similarly, it is shown  that scalar hair $\varphi(r)$ decreases as $\ln r$ increases.
Similarly, for exponential coupling, we  obtain a SCBH  solution for $n = 0$ branch [see Fig.~\ref{fig:5}]. 

%%%%%%%%%%%%%%%%%%%%%%%%%%%%%%%%%%%%%%%%%%%%%%%%%%%%%%%%%%%%%%
\section{Stability of scalarized black holes}\label{5s}

Now, we are in a position to analyze the stability of $n = 0$ branch of SCBHs.
For this purpose, we choose three magnetic charges: $g = 0.175$, $0.200$, and $0.250$ with corresponding bifurcation points given by $\alpha_{n=0} = \{17.338, 12.712, 7.251\}$, respectively.
We  consider two coupling forms: $\tilde{f}(\varphi) = 1 - \alpha \varphi^2$ and $\tilde{f}(\varphi) = e^{-\alpha \varphi^2}$.

Firstly, we introduce radial (spherically symmetric) perturbations around the SCBHs as
\begin{eqnarray}
&&ds_{\text{RP}}^2=-N(r)e^{-2\delta(r)}[1+\epsilon H_0(t,r)]dt^2+\frac{dr^2}{N(r)[1+\epsilon H_1(t,r)]}
+r^2\left(d\theta^2+\sin^2\theta d\varphi^2\right),\nonumber\\
&& \varphi(t,r)=\varphi(r)+\epsilon\frac{ \delta\varphi(t,r)}{r}, 
\end{eqnarray}
where $\varphi(r)$, $N(r)$, and $\delta(r)$ represent the background SCBH  solution, and $H_0(t,r)$, $H_1(t,r)$, and $\delta\varphi(t,r)$ represent the perturbations about it.
We do not need to  introduce a perturbation for gauge field $A_\phi$. Here, $\epsilon$ ($\epsilon \ll 1$) is a control parameter for the perturbations.
From now on, we focus on analyzing the $l = 0$ (s-mode) propagation,neglecting all higher angular momentum modes ($l \neq 0$). In this case, all perturbed fields except the scalar field $ \delta\varphi$ may be considered redundant.

Considering the separation of variables
\begin{eqnarray} \label{Sep-var}
\delta\varphi(t, r) = \varphi_1(r) e^{\Omega t},
\end{eqnarray}
we derive the Schrödinger-type equation for scalar perturbations as
\begin{eqnarray} \label{Schrödinger-type}
\frac{d^2 \varphi_1(r)}{dr_{*}^2} - \left[\Omega^2 + V_{\text{SCBH}}(r)\right] \varphi_1(r) = 0,
\end{eqnarray}
where $r_*$ is the tortoise coordinate defined by
$\frac{dr_*}{dr} = \frac{e^{\delta(r)}}{N(r)}$,
and its potential reads as
\begin{align}
\label{VRwdx1}
V_{\text{SCBH}}(r) &= \frac{e^{-2 \delta(r)} N(r) }{
r^2 \left( g^2 + r^2 \right)^{5/2}}
\left[ \left( g^2 + r^2 \right)^{5/2} - 6 g^2 M r^2 \tilde{f}(\varphi) - \left( g^2 + r^2 \right)^{5/2} N(r)  + 12 g^2 M r^3 \tilde{f}'(\varphi) \varphi'(r) \right. \nonumber\\
&\quad \left. - 2 r^2 \left( g^2 + r^2 \right)^{5/2} \varphi'(r)^2 + 12 g^2 M r^4 \tilde{f}(\varphi) \varphi'(r)^2 + 3 g^2 M r^2 \tilde{f}''(\varphi) \right]
\end{align}

\begin{figure}[H]
\centering
\subfigure[Quadratic coupling: $\tilde{f}(\varphi) = 1 - \alpha \varphi^2$ ]{
\label{fig:6}
\includegraphics[width=7cm]{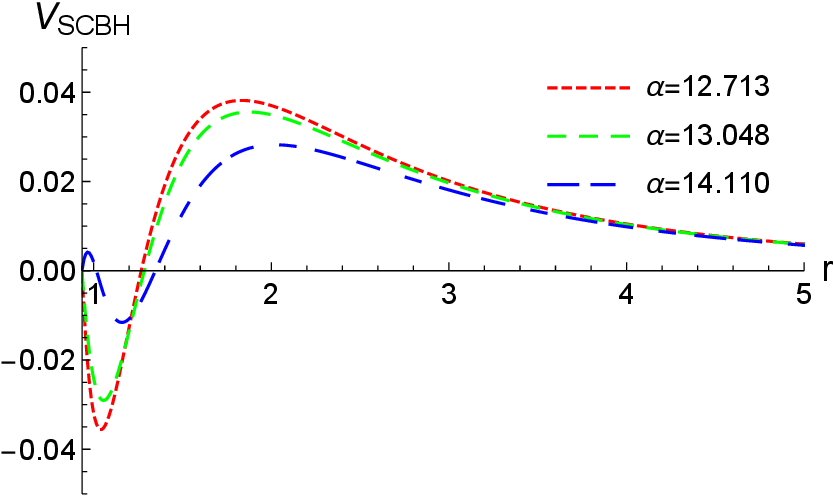}}
\hfill%
\subfigure[Exponential coupling: $\tilde{f}(\varphi) = e^{-\alpha \varphi^2}$]{
\label{fig:8}
\includegraphics[width=7cm]{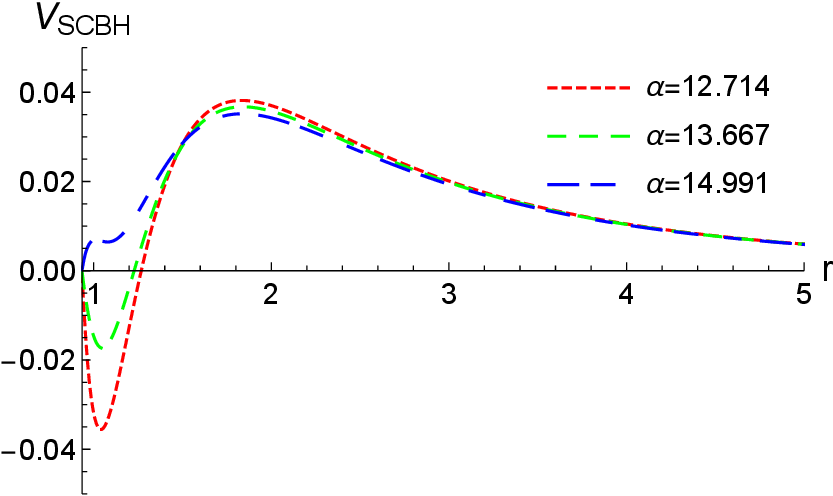}}
  \caption{Three scalar potentials $V_{\text{SCBH}}$  for  $l = 0$ scalar mode around the $n = 0$ branch. Even though they contain small negative regions in the near-horizon, these turn out to be  stable black holes.}
\end{figure}

For quadratic coupling, as suggested by Fig.~\ref{fig:6}, the potentials around the $n = 0$ branch  show small negative regions in the near-horizon, which may indicate instability.
However, a small negative region in the potential $V_{\text{SCBH}}$ with $\alpha = 12.713$ (or $g = 0.2$) does not necessarily imply instability and may instead indicate stability.
The linearized scalar equation (\ref{Schrödinger-type}) around the $n = 0$ branch may support either a stable (decaying) mode with $\Omega < 0$ or an unstable (growing) mode with $\Omega > 0$.

\begin{figure}[H]
\centering
\subfigure[Quadratic coupling: $\tilde{f}(\varphi) = 1-\alpha \varphi^2$ ]{
\label{fig:7}
\includegraphics[width=7cm]{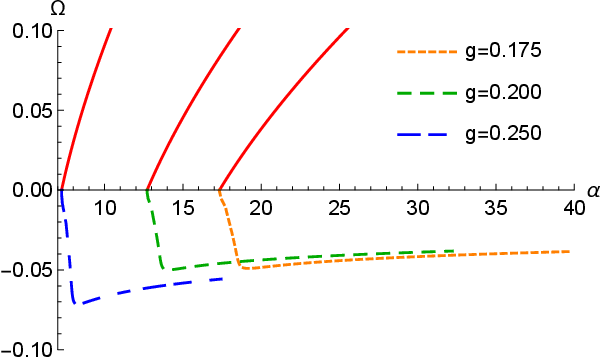}}
\hfill%
\subfigure[Exponential coupling: $\tilde{f}(\varphi) = e^{-\alpha \varphi^2}$]{
\label{fig:9}
  \includegraphics[width=7cm]{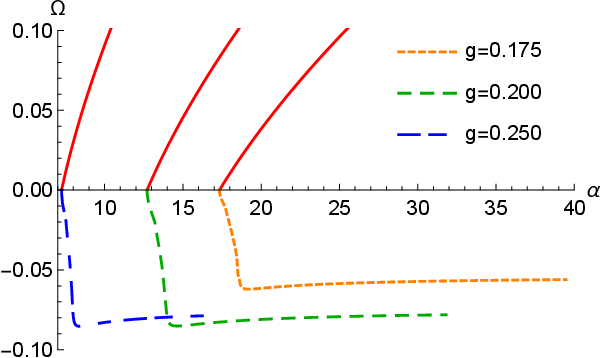}}
  \caption{The negative $\Omega$ is given as a function of $\alpha$ for the $l = 0$ scalar mode around the $n = 0$ branch, showing stability. Here we consider three different cases of $g = 0.175$, $0.200$, and $0.250$. Three dotted curves start from $\alpha_{n=0} = 17.338$, $12.712$, and $7.251$. Three red lines denote the unstable Bardeen black holes [see Fig.~\ref{fig:2}].}
\end{figure}

To fix it, we have to solve Eq.~(\ref{Schrödinger-type}) numerically with vanishing $\varphi_1(r)$  at the horizon and infinity.
From Fig.~\ref{fig:7}, we find that the $n = 0$ black hole is stable against the $l = 0$ scalar mode. Additionally, we show that the stability (or instability) of $n = 0$ black holes is independent of the magnetic charge $g$.

%\subsection{exponential coupling}
%We also perform stability analyses by repeating the procedure in Sec.~\ref{5s-1}, after replacing $1 - \alpha \varphi^2$ with exponential coupling $e^{-\alpha \varphi^2}$. 
For exponential coupling, we also obtain the potential $V_{\text{SCBH}}$ for $n = 0$ branch  (see Fig.~\ref{fig:8}), which is very similar to the potentials shown in Fig.~\ref{fig:6}. The $n = 0$ branch exhibits a large positive region outside the horizon, suggesting stability.

To determine the stability or instability of scalarized black holes, we need to solve the exponential version of equation (\ref{Schrödinger-type}) numerically. This is done by imposing the boundary condition that the redefined scalar field $\tilde{\varphi}_1(r)$ has an outgoing wave at infinity and an ingoing wave at the horizon.
From Fig.~\ref{fig:9}, we find that the $n = 0$ black hole is stable against the $l = 0$ scalar mode because its $\Omega$ is negative. 
This indicates that introducing the exponential coupling does not affect the stability of scalarized Bardeen black holes.

%\begin{figure}[H]
%  \centering
%  \includegraphics[width=8cm]{r-VSBH-e.eps}%\\
%  \caption{Three scalar potentials $V_{\text{SBH}}$ with %$l = 0$ scalar mode, around the $n = 0$ branch with exponential coupling, show stable black holes despite containing small negative regions outside the horizon.}
%  \label{fig:8}
%\end{figure}

%\begin{figure}[H]
%  \centering
%  \includegraphics[width=8cm]{QNMS-alpha-omega-e.eps}%\\
%  \caption{The negative $\Omega$ is given as a function of $\alpha$ for the $l = 0$ scalar mode around the $n = 0$ branch with exponential coupling, showing stability. Here, we consider three different cases of $g = 0.175$, $0.200$, and $0.250$. Three dotted curves start from $\alpha_{n=0} = 17.338$, $12.712$, and $7.251$. Three red lines denote the unstable Bardeen black holes [see Fig.~\ref{fig:2}].}
 % \label{fig:9}
%\end{figure}

\section{Discussions}\label{6s}

In this work, we investigated the spontaneous scalarization of Bardeen black holes.
The computational process is as follows: detecting tachyonic instability of Bardeen black holes $\rightarrow$ predicting scalarized Bardeen black holes (bifurcation points) $\rightarrow$ obtaining the $n = 0$ branch of SCBHs with both quadratic and exponential couplings $\rightarrow$ performing the (in)stability analysis of this branch.

Firstly, we note that the Bardeen black hole is unstable for $\alpha > \alpha_{n=0}(g)$ [see Figs.~\ref{fig:7} and \ref{fig:9}], while it is stable for $\alpha < \alpha_{n=0}(g)$. Here, $\alpha_{n=0}(g)$ denotes the threshold of instability for the Bardeen black hole and indicates the boundary between Bardeen and $n = 0$ branch.
Consequently, the $n = 0$ branch  can be found for any $\alpha \geq \alpha_{n=0}(g)$ with both quadratic and exponential couplings. We also find that the bifurcation point $\alpha_{n=0}(g)$ increases as $g$ decreases. Therefore, the tachyonic  instability becomes harder to realize for smaller magnetic charges. We expect to have infinite ($n=0,1,2,\cdots$) branches of SCBHs because  all SCBHs are found by spontaneous scalarization. All other branches ($n\not=0$) seem to be unstable against radial perturbations as suggested by Refs.~\cite{Myung:2018vug,Myung:2018jvi}. 

Finally, we have shown that the $n = 0$ branch of SCBHs, obtained with both quadratic and exponential couplings, are stable against radial perturbations.
Since the $n = 0$ branch of SCBHs is stable, it is considered as an end point of the Bardeen black hole. Hence, observational implications of this branch are possible to occur~\cite{Stuchlik:2019}.

% \vspace{2cm}

{\bf Acknowledgements}

Q. Y. Pan is supported by National Natural Science Foundation of China (Grant Nos. 12275079 and 12035005).
D. C. Zou is supported by National Natural Science Foundation of China  (Grant No. 12365009) and Natural Science Foundation of Jiangxi Province (No. 20232BAB201039).

\end{document}